\documentclass[twocolumn,amsmath,floatfix,amssymb,aps,prl,showpacs, superscriptaddress]{revtex4-1}
\usepackage{graphicx}

\usepackage{amssymb,amsmath}
\usepackage{dcolumn}
\usepackage{bm}
\usepackage{color,soul}
\usepackage{hyperref}

\begin{document}
\pacs{75.25.-j, 75.70.Rf}
\title{Distinct magnetic phase transition at the surface of an antiferromagnet}

\author{S. Langridge}
\email[Email: ]{sean.langridge@stfc.ac.uk}
\affiliation{ISIS Facility, STFC Rutherford Appleton Laboratory, Harwell Science and Innovation Campus, Oxon, OX11 0QX, U.K.}
\author{G.M. Watson}
\altaffiliation[Current Address: ]{Acorn Packet Solutions 5320
Spectrum Drive Frederick, MD 21703, USA}
\affiliation{Brookhaven National Laboratory, Upton New York 11973-5000, USA}
\author{D. Gibbs}
\affiliation{Brookhaven National Laboratory, Upton New York 11973-5000, USA}
\author{J. J. Betouras}
\affiliation{Department of Physics, University of Loughborough, Loughborough LE11 3TU,
U.K.}
\author{N. I. Gidopoulos}
\affiliation{Department of Physics, Durham University, South Road, Durham, DH1 3LE, U.K.}
\author{F. Pollmann}
\affiliation{Max Planck Institute for Physics of Complex Systems, Noethnitzer Strasse 38, 01187 Dresden,
Germany}
\author{M.W. Long}
\affiliation{School of Physics, Birmingham University, Edgbaston, Birmingham, B15 2TT, U.K.}
\author{C. Vettier}
\affiliation{European Synchrotron Radiation Facility, BP 220 F-38043 Grenoble Cedex, France}
\author{G.H. Lander}
\affiliation{European Commission, Joint Research Center, Institute for Transuranium Elements, Postfach 2340, D-76125, Karlsruhe, Germany}


\date{\today}

\begin{abstract}
In the majority of magnetic systems the surface is required to order at the same temperature as the bulk. In the present study, we report a distinct and unexpected surface
magnetic phase transition, uniquely at a lower temperature than the N\'eel temperature. Employing grazing incidence X-ray resonant magnetic scattering we have observed the near surface behavior of uranium dioxide. UO$_2$ is a non-collinear, triple-{\bf q},
antiferromagnet with the U ions on an face-centered-cubic lattice. Theoretical investigations establish that at the surface the energy increase, due to the lost bonds,
reduces when the spins near the surface rotate, gradually losing  their
normal to the surface component.
At the surface the lowest-energy spin configuration has a double-{\bf q} (planar) structure.
With increasing temperature, thermal fluctuations saturate the in-plane crystal field anisotropy at the surface,
leading to soft excitations that have ferromagnetic $XY$ character and are decoupled from the bulk.
The structure factor of a finite two-dimensional $XY$ model, fits the experimental data well for several orders of magnitude of the scattered intensity.
Our results support a distinct magnetic transition at the surface in the Kosterlitz-Thouless universality class.
\end{abstract}

\maketitle

An interesting question is how the magnetism at the surface of a magnet differs from its bulk magnetism. Understanding the r\^{o}le of the breaking of translational and inversion symmetry and how the bulk structure and order terminate at the surface of materials is the subject of the rich field of electronic reconstruction \cite{Okamoto2004}. Experimental control of such symmetries leads to emergent behavior with charge, orbital, and spin order not present in the bulk \cite{Hwang2012}. Moreover, for the case of antiferromagnets, it has recently been shown that spintronic effects can be realized affording new opportunities for device applications \cite{Park2011}.

Usually in magnets, the onset of bulk magnetic order coincides with that at the surface having the same transition temperature:
the bulk ordering acts as an effective field for the surface \cite{FisherFerdinand1967,BinderHohenberg1974,RudnickJasnow1982}.
Nevertheless, there are extraordinary magnets with stronger magnetic interactions on the surface,
which can promote order at a higher temperature than the bulk \cite{LubenskyRubin, BrayMoore1977}.
However, given the coupling between the bulk and the surface, surface antiferromagnetic (AFM) ordering at a critical temperature below the N\'eel temperature of the bulk has not been reported nor predicted yet. Our study provides the first observation of this situation as well as a theoretical description.

Although surface magnetism has been investigated theoretically for many years \cite{Binder1986}, there are still relatively few experimental scattering studies. This is in contrast to structural studies which are well developed. The emphasis on structural transitions is not surprising given the difficulty in studying the magnetic order from effectively hundreds of pico-grams of
material. Only with the photon brightness of a third generation x-ray source and the amplified sensitivity to
magnetism afforded by resonant x-ray scattering (RXS) \cite{hannon} is this feasible.
In this work, we utilize both of these developments, together with the enhancement in surface sensitivity arising from grazing incidence. Furthermore, we exploit the large resonant enhancement in the magnetic scattering cross-section at the uranium $\rm{M_{IV}}$ absorption edge \cite{uas},
to study the magnetism at the surface of antiferromagnetic UO$_2$.

This dramatic resonant enhancement makes UO$_2$ the best candidate to observe surface magnetism.
In addition, its magnetism gives rise to unexpected behavior close to the surface. For example, at the surface of UO$_2$,
the bulk first order magnetic transition becomes continuous \cite{uo2}.
UO$_2$ has the CaF$_2$ crystal structure, where the uranium, U$^{4+}$,
cations reside on an face centred cubic (fcc) lattice with eight nearest-neighbor oxygen, O$^{2-}$, anions forming a cube.
Below the N{\'e}el temperature T$_N$, the magnetic dipole and the electric quadrupole moments \cite{wilkins2006, rmp2009}
of the two 5\textit{f} electrons on the U$^{4+}$ cations adopt long range AFM order of the transverse triple-{\bf q} type.

The triple-{\bf q} AFM order results from the balancing of three single-{\bf q} components of AFM order:
(001)-type planes with ferromagnetic and electric ferro-quadrupolar order are stacked antiferromagnetically along
the three equivalent $<001>$ directions.
At the same time, the oxygen anion cage that surrounds each U$^{4+}$ cation distorts in such way that cubic symmetry is preserved.
At T$_N \sim  30.2$ K, there is a discontinuous order-disorder transition and for $T > T_N$ the long range magnetic order, the electric quadrupole order and the Jahn-Teller distortion of the oxygen cage, all disappear \cite{wilkins2006}.

\begin{figure}[htb]
\begin{center}
\includegraphics[scale=0.5]{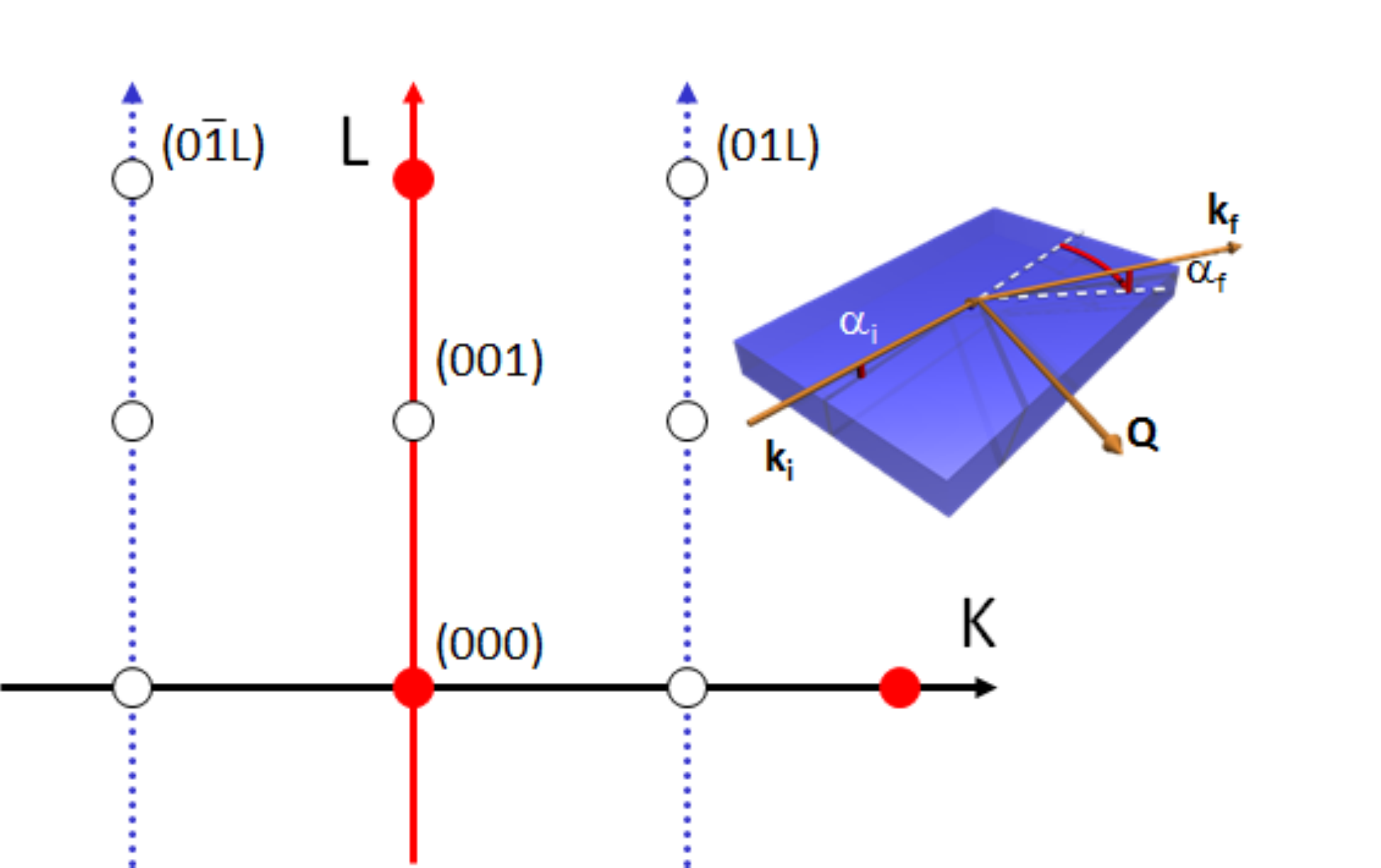}
\end{center}
\caption{fcc allowed structural Bragg reflections in the $(001) \times (010)$ plane: H, K, L are all even or all odd. The structural or charge bulk Bragg spots are solid red. The magnetic Bragg spots are open. The red solid lines show TRs which are both structural and magnetic. The blue dotted lines are pure MTRs. The inset schematically depicts the grazing incidence scattering geometry employed in the experiment. $\alpha_{i(f)}, \textrm{\textbf{k}}_{i(f)}$ are the incident(final) angle and wavevector respectively. $\textrm{\textbf{Q}}$ is the wavevector transfer which is mainly contained within the surface.
}
\label{fn1}
\end{figure}

A key technique to provide atomic resolution information on surface ordering is grazing incidence X-ray scattering \cite{Feidenhans'l1989}.
The abrupt termination of a crystal at its surface gives rise to rods of diffuse scattering, the so-called structural or charge truncation rods (CTRs), which are parallel to the surface normal and connect the charge Bragg reflections \cite{uo2}.
The variation of the scattered intensity along the CTRs gives information about the electronic charge density near the surface. Analogously, the abrupt truncation of magnetic order at the sample surface gives rise to magnetic truncation rods (MTRs) as shown in Fig. \ref{fn1}.
In this work, we report measurements of MTR scattering that allows us to obtain information about the magnetic configuration near the surface of UO$_2$.


For the experimental geometry, the RXS signal is sensitive to the component of the uranium magnetic moment parallel to
the scattered wavevector ($\textbf{k}_f$), which lies primarily in the surface (See inset Fig.~\ref{fn1}).
The chosen position on the MTR for our measurements is correlated to the depth accessed by our probe;
giving a near-surface sensitivity of about 10 nm,
measurements at Bragg spots are essentially bulk whereas away from the Bragg spots the measurements are dominated by surface scattering.
A characterization of the temperature dependence of the MTR will therefore allow us to quantify the change of magnetic
behavior at the surface.

\begin{figure}[htb]
\begin{center}
\includegraphics[scale=0.5]{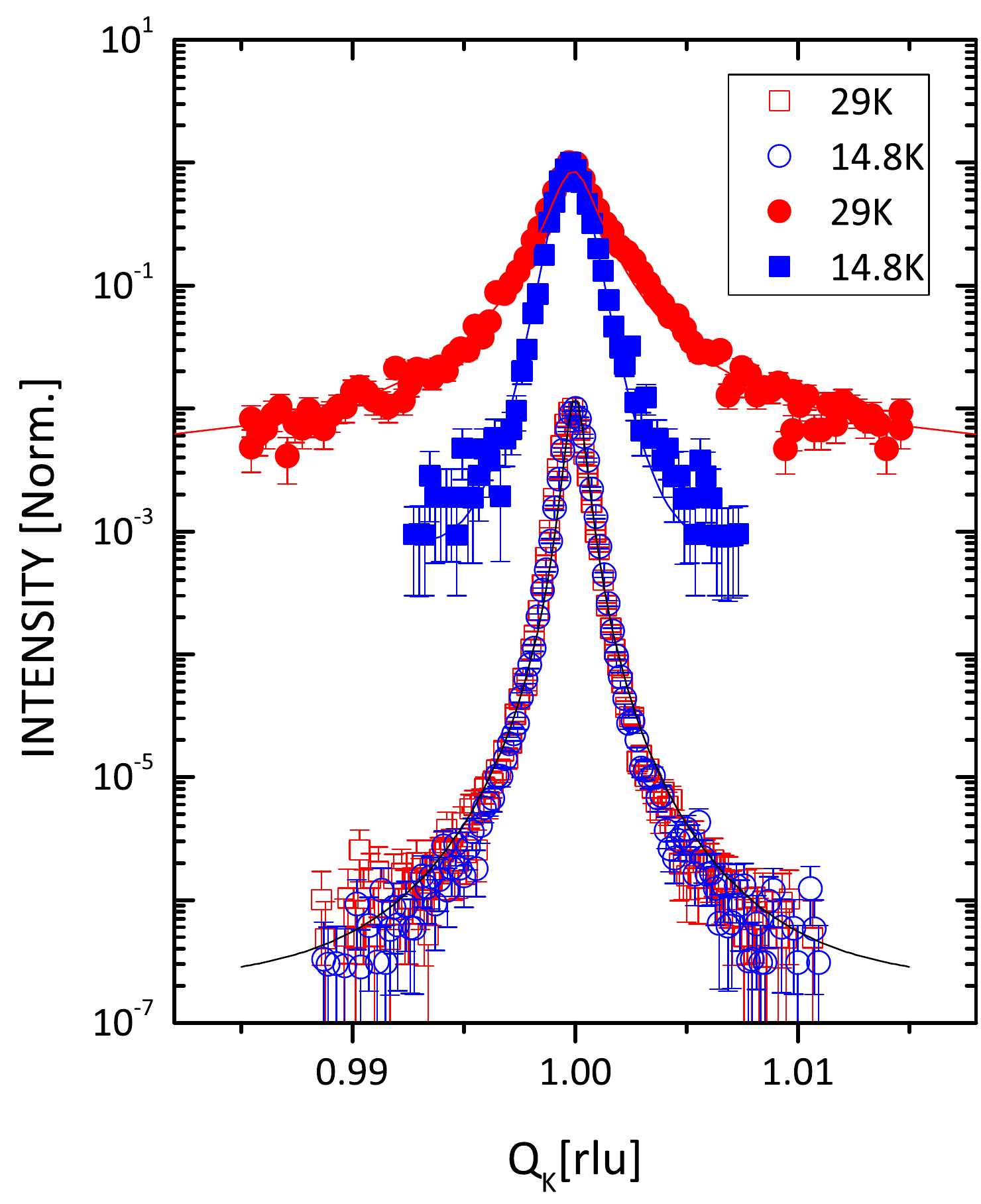}
\end{center}
\caption{Upper curves: The observed normalized intensity (closed symbols) at the magnetic truncation rod $(0 \ Q_K \ 0.97)$ for two temperatures below the bulk N\'{e}el temperature.
Lower curves: The equivalent data (open symbols) for the bulk magnetic Bragg peak $(011)$. The respective data are normalized
and offset for clarity. The solid lines are best fits to the model described in the text.}
\label{fn2}
\end{figure}

A transverse cut through the bulk magnetic $(011)$ Bragg reflection (open symbols) is shown in Fig.~\ref{fn2} (lower curves) for two temperatures below T$_N$.
There is no change in the observed lineshape, which is empirically well described by a Lorentzian function raised to the power of 1.75.
Below the bulk ordering transition (T$_N$), we expect the bulk Bragg reflection lineshape to be temperature independent,
as observed.
In contrast, the magnetic scattering (Fig.~\ref{fn2} upper curves) across the $(0 \ Q_K \ 0.97)$ MTR (purely magnetic rod) exhibits a pronounced change
in lineshape with changing temperature.

The initial working hypothesis is two-dimensional (2D) behavior at the surface region.
Indeed, phenomenologically the best fit of the MTR scattering lineshape
is provided by the structure factor $S({\bf Q})$ derived for a 2D
$XY$ model, with a Lorentzian-square distribution of domain sizes. To obtain this structure factor, we followed the
derivation by Dutta and Sinha \cite{dutta} but used the static spin-spin correlation function
$\langle {\bf S} ({\bf r}) \cdot {\bf S}(0) \rangle $ \cite{Hill1996} in place of the atomic
form factor. For large distances the 2D correlation function decays with distance $r$ as $(r/\xi)^{-\eta}$, where
$\xi$ is the size of the system, here the in-plane, domain size. The surface domain size, $\xi$ below $T_{N}$, where bulk magnetic order is present, is expected to be independent of temperature.

The strength of $\eta \ge 0$ controls how fast spin correlations decay with distance and thus,
it is a measure of the disorder of the magnetism at the surface:
$\eta = 0 $ corresponds to perfect order, while finite values of $\eta > 0$ correspond to quasi-long range order.
The explicit form of $S({\bf Q})$ is given in the supplementary material.
The in-plane domain size at base temperature (15 K) is found to be $\xi=250\pm10\rm{nm}$ and is consistent with that found in Ref.~\cite{uo2}.

\begin{figure}[htb]
\begin{center}
\includegraphics[scale=0.50]{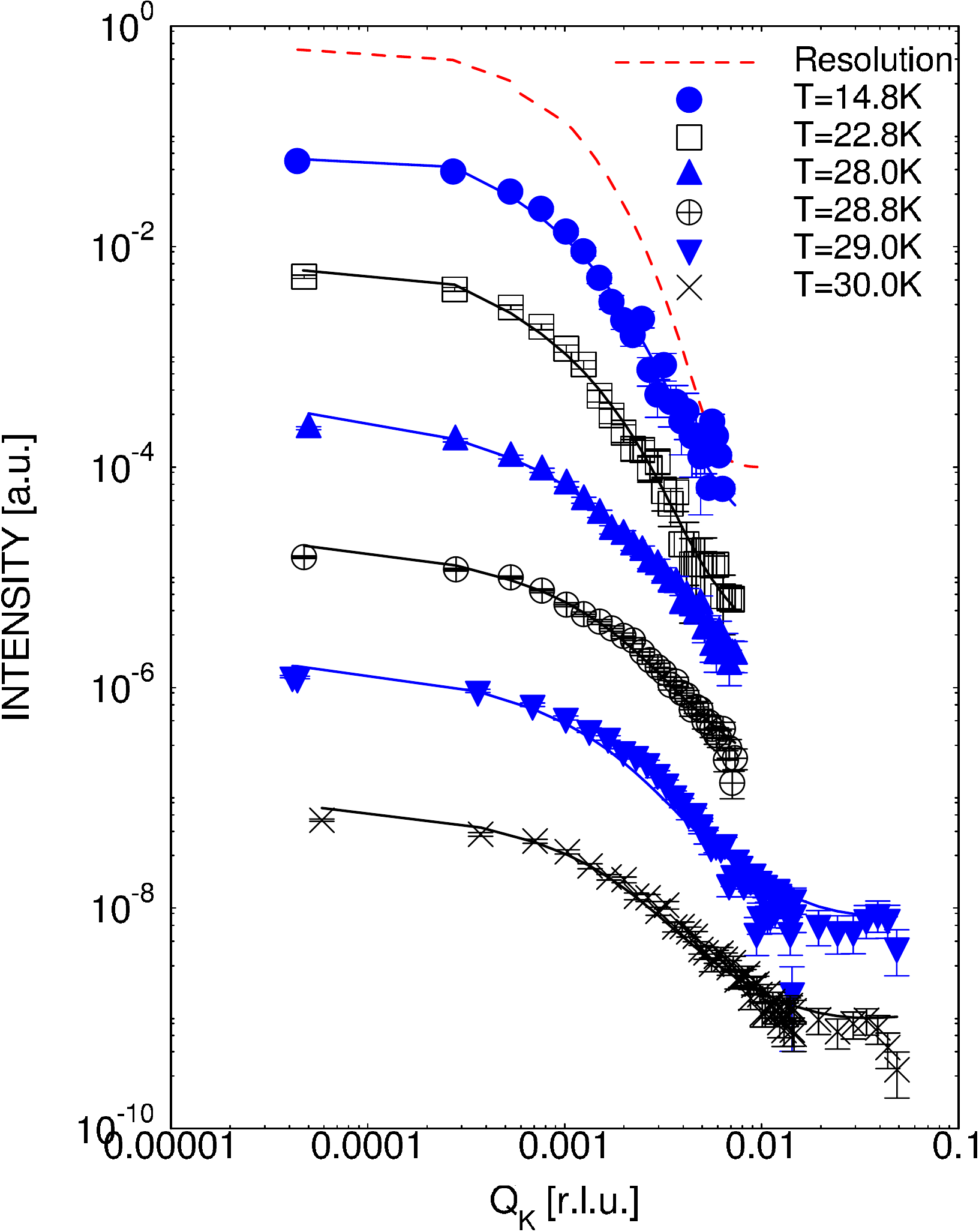}
\end{center}
\caption{
Intensities versus wave-vector as a function of temperature.
One can observe that for temperatures above 22.8~K the reduction in intensity as a function of $Q_K$ is much more gradual (linear) than for lower temperatures as shown explicitly for two temperatures in Fig. 2. This is a key signature of magnetic roughening. The data are offset for clarity.
}
\label{fn3}
\end{figure}

In Fig. \ref{fn3} we show MTR intensities \textit{vs.} $Q_K$ collected at several temperatures below T$_N$. We also show their fits,
with $\eta$ and the amplitide as the only parameters, using the 2D structure factor $S({\bf Q})$. The quality of the fit is excellent
over several orders of magnitude in intensity and all temperatures measured. The dotted (red) line corresponds to $\eta = 0$
and gives the convolution of the delta function $\delta(Q)$ with the experimental resolution.

\begin{figure}[htb]
\begin{center}
\includegraphics[scale=0.50]{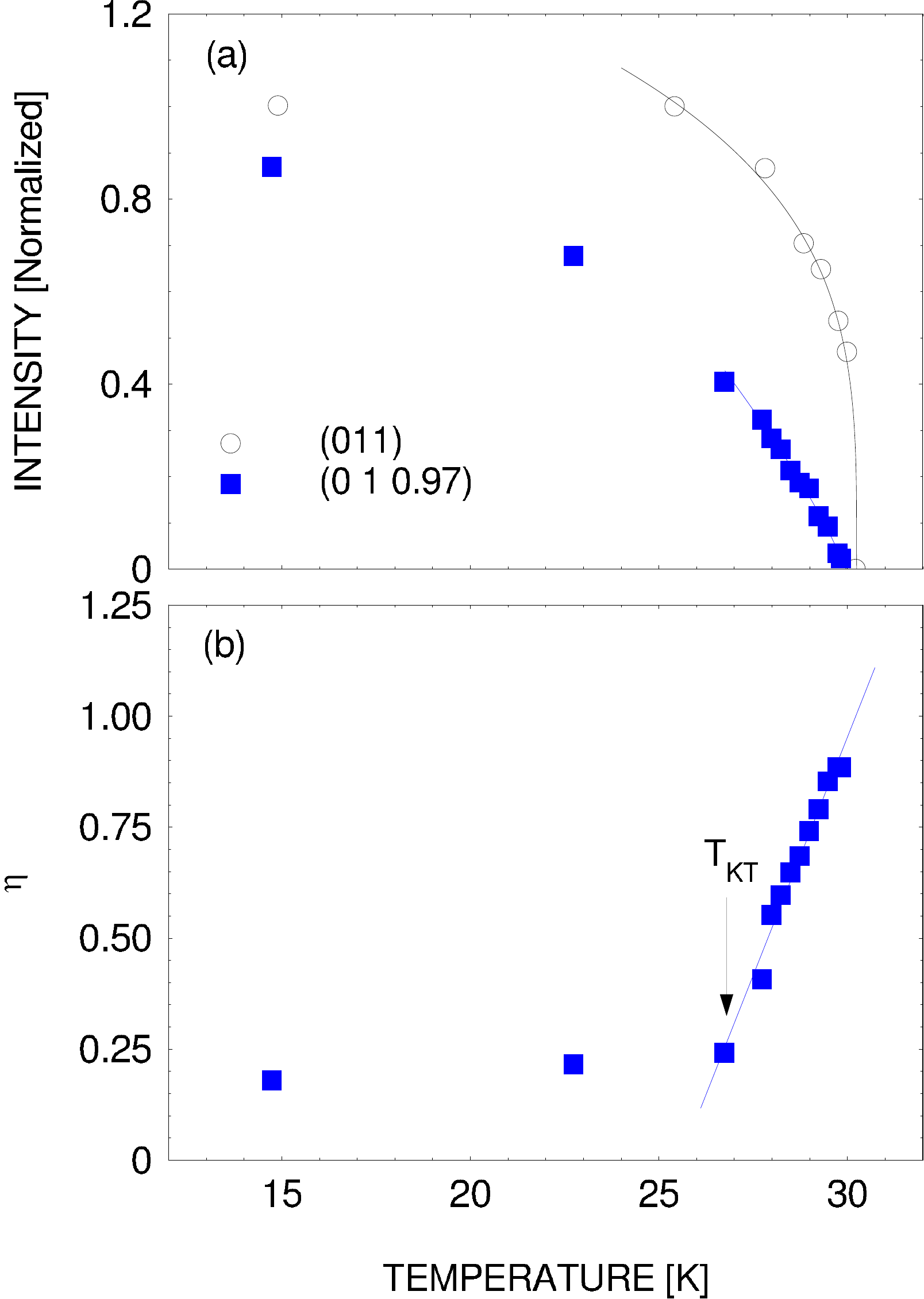}
\end{center}
\caption{ (a)
Open circles show the integrated intensity of the (011) magnetic Bragg reflection as a function of temperature. The transition
is first order (discontinuous). The line connecting the points is a fit to the expression
$ I_0 (1 - {T \over T_N} )^{2 \beta} $ with $\beta = 0.3$.
The blue squares show the integrated intensity at the MTR (0 1 0.97) near the (011) Bragg spot.
The transition at the surface appears continuous.
(b) The variation of $\eta$ as a function of temperature. A discontinuity is identified at a temperature, $T_{KT}$.}
\label{fn4}
\end{figure}

Finally, in Fig. \ref{fn4}a we confirm, with new data, a striking observation first reported in
Ref. \cite{uo2}: the bulk first order transition at the surface appears continuous.
In Fig. \ref{fn4}b we show the fitted values of $\eta$ as a function of temperature. A discontinuity is observed at a new characteristic temperature $T_{KT}=26.8K$. Looking forward we shall discuss this transition in the context of the Kosterlitz-Thouless (KT) transition~\cite{kt,berezinskii}. For temperatures below $T_{KT}$ the exponent $\eta (T)$ is almost constant, while above $T_{KT}$, the exponent $\eta $ increases
rapidly. The critical value, $\eta (T_{KT}) = 0.25$, is consistent with a Kosterlitz-Thouless transition at the
surface \cite{kt,berezinskii}.

The KT transition, in two dimensions, is characterized by the appearance of algebraic order
in the system; it is associated with vortex unbinding, in the original theoretical treatment of the $XY$
model \cite{kt,berezinskii}.
This type of transition is observed in structural surface roughening, which we now briefly discuss, although our magnetic
transition corresponds directly to the original theoretical formulation.

The surface roughening transitions and the KT transition are in the same universality class. Below the roughening transition
temperature the surface is smooth, while above the surface height is no longer well defined. In a
scattering experiment, below the critical temperature, a measurement of the transverse lineshape of the CTR should yield a
$\delta$ function (ignoring resolution effects). Above the critical temperature, the scattering is described by a power
law line-shape characterized by the same exponent, $\eta$, and giving rise to power law singularities:
$S(q)\propto q^{\eta - 2}$.
This exponent $\eta$ is then a measure of the surface roughness. Surface induced order-disorder transitions~\cite{lipowsky}
have been observed in a range of systems, including the chemical surfaces of Ag(001)~\cite{held},
$\rm{Cu}_{3}\rm{Au}$~\cite{cu3au}.
In addition, KT transitions have been also observed recently
in trapped atomic gases \cite{hadzibabic}, in exciton-polariton condensates \cite{roumpos}, and in a photonic lattice \cite{situ_2013}. Even though it was suggested as the archetypal example, an observation in a purely magnetic system is extremely rare and this is the first time that it takes place at the surface of a three-dimensional magnet.

For a KT transition the basic ingredients are the two spatial dimensions and an $XY$ behavior of the magnetic moments.
Therefore the main questions to address the explanation of the present results are:
(i) whether the surface of UO$_2$ is decoupled from the bulk, demonstrating 2D behavior and (ii)
whether the surface of UO$_2$ is then described by an $XY$ Hamiltonian.

In UO$_2$ and for temperatures below T$_N$ there is simultaneously magnetic \cite{wilkins2006,rmp2009}
and electric quadrupolar triple-{\bf q} AFM order, and a Jahn-Teller  distortion of the oxygen cage surrounding the U$^{4+}$  cations,
that retains its cubic symmetry following the triple-{\bf q} bulk order of the U$^{4+}$ cations.
In the following analysis we subsume into a generic anisotropy a potentially essential
ingredient, namely the Jahn-Teller distortion of the oxygen cage.

The unusual and sophisticated transverse triple-{\bf q} magnetic order only occurs
in geometrically frustrated geometries, such as UO$_2$, with the uranium atoms sitting on a fcc lattice
\cite{rmp2009}.
The main sophistication is that there are three styles of long-range order,
and each of these states are simultaneously present in equal amounts in a triple-{\bf q} magnet.

The triple-{\bf q} states, have several unusual characteristics. Firstly, due to the non-collinear spins, there is a
huge magnetoelastic
coupling which leads to hybridisation between phonons and spin-waves~\cite{Caciuffo1999,rmp2009}. Secondly,
and more pertinent,
the spins can heal disorder~\cite{martin}: neighboring non-collinear spins can rotate a component parallel to each-other locally and
the ordering can be subtly altered in such a way that usual magnets cannot. Indeed, the low energy spin-wave corresponds to
precisely such distortions which are energetically the lowest energy fluctuations of the magnet.

The relevant Hamiltonian of a triple-{\bf q} magnetic structure has been discussed previously \cite{perbak}. It includes a Heisenberg term as well as some anisotropic ones.
The small spin-wave gap, observed in UO$_2$, at the magnetic reciprocal lattice points \cite{caciuffo2011} demonstrates that the Heisenberg
energy dominates and indicates that multiple-{\bf q} deformations are the lowest energy magnetic excitations. The anisotropic terms though,
lead to the stability of a triple-{\bf q} structure.

It is essential to understand how this Hamiltonian is optimized in the presence of a surface.
The surface severely affects one of the three equivalent triple-{\bf q} magnetic structures.
Starting from the triple-{\bf q} state in the bulk it seems natural to eliminate this third component and leave a
double-{\bf q} state at the surface which tolerates the loss of bonding. The energetic expense is the anisotropy energy and
the balance between these two energy-scales determines how the angle $\psi$, between the moments and the cube axes, changes smoothly, between the bulk value
$\psi_\infty = 35.26^{\rm o}$ ($\psi_\infty = \arcsin{(1/\sqrt{3})}$) for the triple-{\bf q} state and the surface
value $\psi_0 = 0^{\rm o}$ for the double-{\bf q} state.
This picture has been confirmed by classical Monte Carlo simulations of finite clusters with periodic boundary conditions in
the xy-plane of the fcc lattice (with area $L_x \times L_y$)
and open boundary conditions in the z-direction (which extends from $-L_z/2$ to $+L_z/2$ so that the plane perpendicular to
the z-axis at $z=0$ represents the bulk of the crystal).
Finite size scaling then confirms that the triple-{\bf q} antiferromagnetism becomes double-{\bf q}  as the surface is
approached. These calculations clearly establish the XY character of the magnetic structure and the absence of a moment perpendicular to the surface. The details of these calculations, which partly contribute to the explanation of a KT transition, are beyond the scope of the present letter and will appear elsewhere.

In UO$_2$, in-plane deviations from the double-{\bf q} surface state are only weakly coupled to the bulk and with nearly isotropic, plane rotor character.
There is a weak residual crystal-field potential, which at low enough temperatures, destroys the isotropy in the $xy$ plane.
Because the surface-bulk coupling does not vanish, thermodynamically, no order that exists in the bulk can disappear sharply below the surface.
Given that all three triple-{\bf q} order parameters are present up to the
first-order transition in the bulk, in principle, none of these orders can control a surface transition.
However, the analogy with similar unusual behavior in the bulk helps to explain our findings.

In Sr$_2$YRuO$_6$, a partial long-range ordered state with coupled alternate AFM YRuO$_4$ square layers coexisting with the short-range correlations is developed below
T$_{N1}$ =32 K and a second transition to a fully ordered AFM state below T$_{N2}$=24 K \cite{Granado}. The reduced
dimensionality of the spin correlations is arguably due to a cancelation of the magnetic coupling between
consecutive AFM square layers in fcc antiferromagnets.
In UO$_2$ a similar behavior with more spectacular results, comes from the almost independence of the surface and the bulk which are weakly coupled.

The puzzle to understand the two bulk phase transitions in these type-I antiferromagnets is the same:
at intermediate temperatures between the two transitions,
the existing order of the higher temperature phase is expected to influence and keep the
lower temperature phase ordered too. The way out is that the symmetries of the two phases are different.
Theoretical studies of the planar rotor model, in the presence of a $p$-fold crystal field anisotropy establish \cite{Jose}
that above a critical temperature thermal fluctuations saturate the anisotropy field, when $p \ge 4$, and reduce the model to an effectively isotropic planar rotor.
At the surface of UO$_2$, the energy scale of the anisotropy field is the same as the coupling with the bulk, and
the excitations of the double-{\bf q} surface state also thermally decouple from the bulk.
Hence, the surface magnetic transition of the UO$_2$ can be interpreted as the broken symmetry, 
double-{\bf q} surface layer pointing along the anisotropy directions, destabilizing into an isotropic,
spatially-varying, power-law controlled, double-{\bf q} state.

The angle that characterizes the style of double-{\bf q} state is
the variable that loses its algebraic order; thus providing the second crucial theoretical ingredient of $XY$ model behavior.
When in-plane excitations of the surface state become thermodynamically isotropic, they also effectively decouple from the bulk.
In the experiment, this is the region where the power law exponent $\eta$ increases rapidly with increasing temperature.

As a consequence, the physical picture of the distinct surface behavior of the
triple-{\bf q} magnetism in UO$_2$ now emerges.
The bulk has a triple-{\bf q} state with a spin-wave gap that softens close to T$_N$.
The expected bulk transition to a spatially varying
multiple-{\bf q}  magnetism is very well characterized as a first order transition.
At the surface, the spins flatten, providing a local
double-{\bf q}  state with weakened anisotropy.
This state does indeed suffer a continuous transition to a 2D, spatially varying
double-{\bf q}  state, 
at a different characteristic temperature.

The observed remarkable effect of the surface, seemingly acting differently from the bulk,  is a consequence of the
magnetic frustration; this can lead to
a two-step process towards final order, both in the bulk and independently on the surface. This phenomenon has been
revealed as a result of the advancement  of the x-ray magnetic scattering technique at third generation synchrotron sources. By producing epitaxial nanoscale thin films of UO$_2$ it should be possible to induce finite thickness effects thereby stabilizing uniquely the double-{\bf q} state which would then be accessible by the methodology outlined in this work. The sharp changes in the correlation functions and the subsequent analysis is consistent with the scenario of a surface
phase transition in the universality
class of Kosterlitz-Thouless, the ramifications of which, should be further investigated theoretically.

We thank the European Synchrotron Radiation Facility for the
provision of beamtime. We gratefully acknowledge useful
discussions with N. Bernhoeft. One of us (SL) gratefully
acknowledges fruitful discussions with M.K. Sanyal, S.K. Sinha and
R.M. Dalgliesh. SL and NG gratefully acknowledge discussions with
Paola Verrucchi on the nature of the KT transition. JJB gratefully
acknowledges  interesting discussions with Jeroen van den Brink and Daniel Khomskii and support from EPSRC under
EP/H049797/1. Work performed at Brookhaven National Laboratory is supported by the US DOE under Contract No.
DE-AC02-CH7600016.

\end{document}


\section{Supplemental Material}

The bulk magnetic characteristics for $\rm{UO}_{2}$ have been known
for over 40 years~\cite{rmp2009}. The allowed Bragg reflections then
correspond to the selection rules $H$, $K$, $L$ (all even or odd)
and the magnetic Bragg reflections are separated from the allowed
chemical Bragg reflections by the reciprocal lattice vector $<001>$.
The reciprocal lattice is as shown in figure~1 of the main letter.

$\rm{UO}_{2}$ has a face centred cubic $\rm{CaF_{2}}$ structure at
room temperature with a lattice parameter of 5.47\AA. $\rm{UO}_{2}$
exhibits a bulk discontinuous phase transition at
$\rm{T}_{N}\approx30.2$K~\cite{rmp2009}. To study experimentally the
ordering of the surface requires a smooth, chemically ordered and
stable surface. Being an oxide, $\rm{UO}_{2}$ has a relatively inert
surface and the largest known resonant enhancements (of order
$10^{6}$~\cite{uas}) of the x-ray magnetic scattering observed as the
incident photon energy is tuned to the actinide absorption M-edges.
 The (001) surface of UO$_2$ was
cut and polished and subsequently annealed in an $\rm{Ar/H}_{2}$
atmosphere at $1400^{\circ}\rm{C}$ to ensure stoichiometry. The
observed sample mosaic was better than $0.05^{\circ}$.  The
experiments were performed on the insertion device beamline ID20 at
the ESRF. The sample was mounted in a Be can filled with He exchange
gas in a closed cycle refrigerator. All measurements were performed
at the U $\rm{M}_{IV}$ edge at 3.728 keV. The experiments were
performed in a glancing incidence geometry with the incidence angle
close to the critical angle for UO$_{2}$
($\alpha_{i}\sim0.75^{\circ}$) giving a near-surface sensitivity of
$\approx 10$nm.

The sample was the same single crystal used in earlier work [4]. Between measurements the crystal, in its Be He-filled can, was kept in an inert atmosphere glove box at the Institute for Transuranium Elements, Karlsruhe. Before measurements for the present work, a check was made at room temperature of the reflectivity and the intensity of the charge-truncation rods. Surface sensitive diffraction clearly shows that these are rapidly affected if the sample surface is contaminated. See, for example, the charge-truncation rods and their changes with additional oxidation of the surface as shown in Fig. 2[4].

As is well understood, the simple power law form for the static
structure factor of an infinite 2 dimensional lattice is modified
by finite size effects. This was analytically described by Dutta
and Sinha~\cite{dutta} for the case of a two-dimensional harmonic
lattice. One may use this as the basis to for an understanding of
the magnetic surface. The most important step is to replace the
atomic form factor $f_{\vec{q}}(\vec{x},0)$ with the static
spin-spin correlation function $G^\prime(\vec{x},0)$ in the
$XY$-model. The domain size distribution is that described by a
Lorentzian squared function resulting in the structure factor:
\begin{equation}\label{sq}
S(\mathbf{q}) \propto
(q\xi)^{-1+\eta}\left[\pi\eta
(q\xi)^{1-\eta}
\frac{_{1}F_{2}\left(1-\frac{\eta}{2};1,\frac{-\eta}{2};\frac{(q\xi)^{2}}{4}\right)}{4sin\left(\frac{\pi\eta}{2}\right)}
\right.
  + \left.\frac{\Gamma(-1-\frac{\eta}{2})(q\xi)^{3}\
  _{1}F_{2}\left(2;2+\frac{\eta}{2},2+\frac{\eta}{2};\frac{q^{2}\xi^{2}}{4}\right)}{2^{3+\eta}\Gamma(2+\frac{\eta}{2})}\right]\\
\end{equation}
where $q$ is the wavevector transfer, $\eta$ the surface exponent,
$_{1}F_{2}$ is the hypergeometric function and $\xi$ is the in-plane
domain size. This function accurately describes the surface Bragg
peaks for $\eta=0$.

The results obtained in the present paper are consistent with those reported in the previous work\cite{uo2}. In that reference a lack of order was noted near the surface at higher temperature (see Fig. 9 of Ref. \onlinecite{uo2}) and there is a strong divergence of the layer thickness of disordered material starting at 27 K (see Fig. 13 of Ref.~\cite{uo2}). Because at that time (at a 2nd-generation synchrotron source) there was insufficient intensity, except at the lowest temperatures when the moments are fully ordered, to perform detailed scans across the magnetic truncation rods, such as shown in the present experiment by Figs. 2 and 3, the exact form of the roughening remained unresolved. As noted in the present paper $T_{KT}$ = 26.8 K, is in excellent agreement with the divergence noted in Ref.~\onlinecite{uo2} (see Fig. 13). The new experiments have revealed the nature of the phase transition at the surface before the material completely disorders.